\documentclass[12pt]{article}
\usepackage{epsfig}
\usepackage{cite}

\begin{document}
\baselineskip 0.7cm

\newcommand{\gsim}{ \mathop{}_{\textstyle \sim}^{\textstyle >} }
\newcommand{\lsim}{ \mathop{}_{\textstyle \sim}^{\textstyle <} }
\newcommand{\vev}[1]{ \left\langle {#1} \right\rangle }
\newcommand{\lsp}{ \left ( }
\newcommand{\rsp}{ \right ) }
\newcommand{\lmp}{ \left \{ }
\newcommand{\rmp}{ \right \} }
\newcommand{\llp}{ \left [ }
\newcommand{\rlp}{ \right ] }
\newcommand{\labs}{ \left | }
\newcommand{\rabs}{ \right | }
\newcommand{\EV} { {\rm eV} }
\newcommand{\KEV}{ {\rm keV} }
\newcommand{\MEV}{ {\rm MeV} }
\newcommand{\GEV}{ {\rm GeV} }
\newcommand{\TEV}{ {\rm TeV} }
\newcommand{\YR}{ {\rm yr} }
\newcommand{\mgut}{M_{GUT}}
\newcommand{\mint}{M_{I}}
\newcommand{\mgra}{M_{3/2}}
\newcommand{\mll}{m_{\tilde{l}L}^{2}}
\newcommand{\mdr}{m_{\tilde{d}R}^{2}}
\newcommand{\mllXX}[1]{m_{\tilde{l}L , {#1}}^{2}}
\newcommand{\mdrXX}[1]{m_{\tilde{d}R , {#1}}^{2}}
\newcommand{\mgy}{m_{G1}}
\newcommand{\mgl}{m_{G2}}
\newcommand{\mgc}{m_{G3}}
\newcommand{\nuR}{\nu_{R}}
\newcommand{\slL}{\tilde{l}_{L}}
\newcommand{\slLi}{\tilde{l}_{Li}}
\newcommand{\sdR}{\tilde{d}_{R}}
\newcommand{\sdRi}{\tilde{d}_{Ri}}
\newcommand{\e}{{\rm e}}
\newcommand{\bsub}{\begin{subequations}}
\newcommand{\esub}{\end{subequations}}
\newcommand{\wt}{\widetilde}
\newcommand{\tm}{\times}
\newcommand{\ra}{\rightarrow}
\newcommand{\del}{\partial}
\newcommand{\az}{a_{Z}^{}}
\newcommand{\bz}{b_{Z}^{}}
\newcommand{\cz}{c_{Z}^{}}
\newcommand{\aw}{a_{W}^{}}
\newcommand{\bw}{b_{W}^{}}
\newcommand{\dw}{d_{W}^{}}
\newcommand{\sw}{s_{W}}
\newcommand{\cw}{c_{W}}
\newcommand{\gz}{g_{Z}^{}}
\newcommand{\mz}{m_{Z}^{}}
\newcommand{\pH}{p_{H}^{}}
\newcommand{\pone}{p_{1}^{}}
\newcommand{\ptwo}{p_{2}^{}}
\newcommand{\pt}{\partial}
\newcommand{\btable}{\begin{table}[htbp]\begin{center}}
\newcommand{\etable}[1]{ \end{tabular}\caption{#1}\end{center}\end{table} }
\newcommand{\vt}{\vspace{3mm}}
\renewcommand{\thefootnote}{\fnsymbol{footnote}}
\setcounter{footnote}{1}

\makeatletter
%
%
%
%
%
\newtoks\@stequation

\def\subequations{\refstepcounter{equation}%
  \edef\@savedequation{\the\c@equation}%
  \@stequation=\expandafter{\theequation}
  \edef\@savedtheequation{\the\@stequation}
  \edef\oldtheequation{\theequation}%
  \setcounter{equation}{0}%
  \def\theequation{\oldtheequation\alph{equation}}}

\def\endsubequations{%
  \ifnum\c@equation < 2 \@warning{Only \the\c@equation\space subequation
    used in equation \@savedequation}\fi
  \setcounter{equation}{\@savedequation}%
  \@stequation=\expandafter{\@savedtheequation}%
  \edef\theequation{\the\@stequation}%
  \global\@ignoretrue}


\def\eqnarray{\stepcounter{equation}\let\@currentlabel\theequation
\global\@eqnswtrue\m@th
\global\@eqcnt\z@\tabskip\@centering\let\\\@eqncr
$$\halign to\displaywidth\bgroup\@eqnsel\hskip\@centering
     $\displaystyle\tabskip\z@{##}$&\global\@eqcnt\@ne
      \hfil$\;{##}\;$\hfil
     &\global\@eqcnt\tw@ $\displaystyle\tabskip\z@{##}$\hfil
   \tabskip\@centering&\llap{##}\tabskip\z@\cr}

\makeatother


\begin{titlepage}

\begin{flushright}
UT-02-25
\end{flushright}

\vskip 0.35cm
\begin{center}
{\large \bf Virtual Black Holes at Linear Colliders}
\vskip 1.2cm
Yosuke Uehara$^{a)}$

\vskip 0.4cm

$^{a)}$ {\it Department of Physics, University of Tokyo, 
         Tokyo 113-0033, Japan}\\
\vskip 1.5cm

\abstract{We propose that future linear colliders can create virtual
black holes even though their energies are below the fundamental scale,
because of the uncertainty principle. These virtual black holes provide
us much information which cannot be obtained in the production of 
black holes at the LHC or other hadron
colliders. We can observe lepton flavor and lepton
(baryon) number violating processes at linear colliders by using virtual 
black holes. And virtual black holes can be used
to the precision measurements of top and W. They can create 
only one top quark or one W boson, which leads to the clean signal
that cannot be obtained in pair-production processes.}

\end{center}
\end{titlepage}

\renewcommand{\thefootnote}{\arabic{footnote}}
\setcounter{footnote}{0}

%
%
%
%

\vt

{\bf Introduction.}

\vt

The possibility of low-energy gravity may enable us to produce
black holes at future colliders, say LHC \cite{LHC} or Tevatron
\cite{Tevatron}. And high-energy cosmic rays may also produce
black holes \cite{Cosmicray}. But their studies can be done
in the semiclassical limit, $M_{BH} \gg M_{D}$, where $M_{D}$
is the true fundamental scale. (See \cite{Uehara} for the latest lower
bound of $M_{D}$.) If the mass of black hole $M_{BH}$ 
approaches $M_{D}$, the quantum gravity drastically affect the production
cross section and nobody can calculate it. Therefore future
linear colliders were not considered as the candidates of black hole
factories due to their low energy.

In this letter, we show that even linear colliders can produce
{\bf virtual} black holes, and the information given by 
their decay is the treasure which cannot be obtained from 
the decay of black holes at hadron colliders.

\vt

{\bf The Setup.}

\vt

We consider virtual black holes produced at future linear colliders.
As an input, we set:
\bsub
\begin{eqnarray}
\sqrt{s}&=&1 \TEV, \label{setupeq1} \\
n&=&7, \\ 
M_{D}&=&1 \TEV, \\
M_{BH}&=&5 \TEV, \label{setupeq4}
\end{eqnarray}
\esub
where $n$ is the number of extra dimensions and 
$M_{BH}$ is the mass of {\bf virtual} black hole.
 If we approach $M_{BH}$ to $M_{D}$, sizable
quantum gravity effects appear and we cannot calculate the cross 
section. But as stated in \cite{Giudice-Rattazzi-Wells}, for 
$M_{S}/g_{S} < M_{BH} < M_{S}/g_{S}^{2}$, where $M_{S}$ is the
string scale and $g_{S}$ is the string coupling constant, 
the cross section is given by the semiclassical one. 
(\ref{setupeq1}-\ref{setupeq4}) satisfies this condition and we can
reliably use it.

\vt

{\bf Production and Decay.}

\vt

In the semiclassical limit $M_{BH} \gg M_{D}$, the $(4+n)$-
dimensional Schwarzschild radius is given by \cite{Myers-Perry}:
\begin{eqnarray}
R_{S} \sim \frac{1}{\sqrt{\pi} M_{D}} \left[ \frac{M_{BH}}{M_{D}} (\frac{8 \Gamma(\frac{n+3}{2})}{n+2}) \right]^{1/(n+1)}.
\end{eqnarray}
And semiclassical reasoning implies that a black hole is produced when
the distance between two particles fall into this Schwarzshird radius,
and the cross section is given by:
\begin{eqnarray}
\sigma \sim \pi R_{S}^{2} = \frac{1}{M_{D}^{2}} \left[ \frac{M_{BH}}{M_{D}} (\frac{8 \Gamma(\frac{n+3}{2})}{n+2}) \right]^{2/(n+1)} \sim 3.21 \frac{1}{M_{D}^{2}}.
\end{eqnarray}
But note that we are now considering $\sqrt{s}=1 \TEV$ linear
collider. Thus we cannot produce {\bf real} black holes anyhow.
The method to evade this problem is to use the uncertainty principle.
During the time scale $\Delta t \sim 1/(M_{BH}/2)$, we can violate
the energy conservation low. For each beam bunches we should apply
the uncertainty principle. This leads to the suppression factor:
\begin{eqnarray}
\left( \frac{\sqrt{s}/2}{M_{BH}/2} \right)^{2} \sim \frac{1}{25}.
\end{eqnarray}
But this is not the end of the story. Voloshin \cite{Voloshin} claimed
that the cross section for production of large black holes is
suppressed by at least a factor $\exp (-I_{E})$, where $I_{E}$ 
is the Gibbons-Hawking action for the black hole.

Taking into account all these results, the production cross section
of {\bf virtual} black holes in the assumption 
(\ref{setupeq1}-\ref{setupeq4}) becomes:
\begin{eqnarray}
\sigma = 3.21 \frac{1}{M_{D}^{2}} \ (\frac{1}{25}) \  \exp(-I_{E}) \ (3.89 \tm 10^{11}) \ = 2.1 \tm 10^{4} \ {\rm fb}.
\end{eqnarray}
We can also calculate the temperature of this black hole $T_{BH}$, 
the mean energies of decayed particles $\vev{E}$ and the average 
multiplicity $\vev{N}$. They are given by:
\bsub
\begin{eqnarray}
T_{BH} &=& \frac{n+1}{4 \pi R_{S}} = 0.64 \ \TEV, \\
\vev{E} &=& 2 T_{BH} = 1.3 \ \TEV, \\
\vev{N} &=& \frac{M_{BH}}{2 T_{BH}} = 3.9.
\end{eqnarray}
\esub
Thus one {\bf virtual} black hole emits about 4 particles.
But note that the energies of their particles are not 1.3 \TEV
since the time scale during the uncertainty principle holds is 
very short, and they become $E \sim \sqrt{s}/4 = 250 \GEV$.

\vt

{\bf New Physics: Lepton Flavor and Lepton (Baryon) Number Violation.}

\vt

From the four decay products of one black hole, we can investigate
many exciting physical results. From now we assume the integrated luminosity
${\cal L}=100 {\rm fb}^{-1}$.

\vt

The first one is the 
{\bf test of Hawking radiation,} which is enabled by the missing
transverse energy carried by neutrinos. But since 
this issue was investigated in \cite{LHC}, we do not consider it
in this letter.

\vt

Next, {\bf we can consider lepton flavor violating processes.}
The processes are described as follows:
\bsub
\begin{eqnarray}
e^{+} e^{-} &\ra& \ {\rm \mathbf{virtual}} \ {\rm black hole} \ra \ e^{\pm}  \ \mu^{\mp} \ (q \ \bar{q}), \\ 
e^{+} e^{-} &\ra &\ {\rm \mathbf{virtual}} \ {\rm black hole} \ra \ e^{\pm}  \ \mu^{\mp} \ (g \ g), \\
e^{+} e^{-} &\ra& \ {\rm \mathbf{virtual}} \ {\rm black hole} \ra \ e^{\pm}  \ \mu^{\mp} \ (l \ \bar{l}), \\
e^{+} e^{-} &\ra& \ {\rm \mathbf{virtual}} \ {\rm black hole} \ra \ e^{\pm}  \ \mu^{\mp} \ (\gamma \ \gamma). 
\end{eqnarray}
\esub
Black holes evaporate into the Standard Model(SM) particles without any
discrimination. The number of these processes is calculated to be $590$.
Since the SM has no processes which mediate lepton flavor violation,
$100 {\rm fb}^{-1}$ operation can prove lepton flavor violation
with the accuracy $1/(\sqrt{590}) \sim 4.1 \%$. 
There are no experimental obstacles except muon tracking. 
JLC study \cite{JLC} showed that electron
calorimeter response is better than $0.3 \%$ for $2 \GEV - 250 \GEV$,
and the separation of two pion clusters make it possible to detect
hadronic jet with energy $E \sim 250 \GEV$. The problem is
muon tracking. The current resolution is about $1$ cm, which enables
us to detect muon momentum only for $p_{\mu}^{} \lsim 100 \GEV$. 
So the detector technology is still to be upgraded.

\vt
 
Third, {\bf we consider lepton and baryon number violating processes.} 

\vt

They are:
\bsub
\begin{eqnarray}
e^{+} e^{-} &\ra& \ {\rm \mathbf{virtual}} \ {\rm black hole} \ra \ (\mu^{-} / e^{-})  \ u \ u \ d, \label{NVeq1} \\
e^{+} e^{-} &\ra& \ {\rm \mathbf{virtual}} \ {\rm black hole} \ra \ (\mu^{+} / e^{+})  \ d \ d \ d. \label{NVeq2} 
\end{eqnarray}
\esub
Their charge-conjugated processes also exist. Here $u$ 
denotes the upper sector of quark doublet and $d$ denotes the
lower sector of that. The number of these processes becomes $650$.
Since the SM prreserves lepton and baryon number,
we can prove lepton (baryon) number violating processes with
the accuracy $1/\sqrt{650} = 3.9 \%$.

\vt

{\bf Presicion Measurement: Single Top, Single W}

\vt

Virtual black holes can also be used to the precision measurements.
For example, consider the following processes.
\bsub
\begin{eqnarray}
e^{+} e^{-} &\ra& \ {\rm \mathbf{virtual}} \ {\rm black hole} \ra \ t  \ \bar{u} \ (q \ \bar{q}), \label{topeq1} \\
e^{+} e^{-} &\ra& \ {\rm \mathbf{virtual}} \ {\rm black hole} \ra \ t  \ \bar{u} \ (g \ g), \label{topeq2} \\
e^{+} e^{-} &\ra& \ {\rm \mathbf{virtual}} \ {\rm black hole} \ra \ t  \ \bar{u} \ (l \ \bar{l}), \label{topeq3} \\
e^{+} e^{-} &\ra& \ {\rm \mathbf{virtual}} \ {\rm black hole} \ra \ t  \ \bar{u} \ (\gamma \ \gamma), \label{topeq4} 
\end{eqnarray}
\esub
with their charge-conjugated ones.
The calculation show that we obtain 1300 single top quarks. 
Top quark immediately decays. 
In order to determine the mass of top quark, the following
decay chain is the best.
\begin{eqnarray}
t \ra W^{+} b \ra (e^{+}/\mu^{+}) (\nu_{e/\mu}) b.
\end{eqnarray}
The difference from usual top-pair production is that only one neutrino
is emitted, and thus the combinatorial background is completely zero.
This means only the statistical and the calorimeter error
dominates the error of top quark mass. 

Here we estimate the error. The calorimeter error of JLC is:
\bsub
\begin{eqnarray}
\frac{\sigma_{E}}{E} &=& \frac{15 \%}{\sqrt{E(\GEV)}} \oplus 1 \% \ {\rm for \ e/\gamma}, \\
\frac{\sigma_{E}}{E} &=& \frac{40 \%}{\sqrt{E(\GEV)}} \oplus 2 \% \ {\rm for \ hadrons}. \\
\end{eqnarray}
\esub
The calorimeter error of b-quark with energy $250 \tm 0.5 \GEV$ 
is $4.1 \%$ and it dominates the error of calorimeter. 
We use leptonic decay mode only and 
${\rm Br}(W \ra \ (e/\mu) \ (\nu_{e/\mu}))=0.21$.
So the statistical error is $1/\sqrt{1300 \tm 0.21}=6.1 \%$.

Therefore the total error becomes $7.3 \%$. 
This is comparable value with the error obtained by Tevatron, $5.1 \%$.
If we combine these results, final error becomes $4.5 \%$.
So virtual black holes can reduce the error of top quark mass,
which means that virtual black holes can be the complement
of the current electroweak presicion measurements.

Next we consider single W production processes. They are:
\bsub
\begin{eqnarray}
e^{+} e^{-} &\ra& \ {\rm \mathbf{virtual}} \ {\rm black hole} \ra \ W^{+}  \ l^{-} \ (q \ \bar{q}), \label{topeq1} \\
e^{+} e^{-} &\ra& \ {\rm \mathbf{virtual}} \ {\rm black hole} \ra \ W^{+}  \ l^{-} \ (g \ g), \label{topeq2} \\
e^{+} e^{-} &\ra& \ {\rm \mathbf{virtual}} \ {\rm black hole} \ra \ W^{+}  \ l^{-} \ (l \ \bar{l}), \label{topeq3} \\
e^{+} e^{-} &\ra& \ {\rm \mathbf{virtual}} \ {\rm black hole} \ra \ W^{+} \ l^{-} \ (\gamma \ \gamma), \label{topeq4} 
\end{eqnarray}
\esub
again with their charge-conjugated ones. The calculation shows that
we can obtain $450$ single W bosons.
Since we know that the initial energy of $W$-boson is about $250 \GEV$,
the mass reconstruction process from $W^{+} \ra l^{+} \nu$ is 
straightforward. The only source of the error is the calorimeter resolution.
It is estimated as $1.4 \%$, and it is suppressed by the number of
single W, and the error becomes $0.066 \%$.
The current error is $0.07 \%$, and thus again virtual black holes
can play a complementary role of current measurements.

\vt

{\bf Summary.}

\vt

In this letter we explored new possibility that linear colliders
can be used as {\bf virtual} black hole production machines.
The cleanness of linear colliders enables us to analyze 
new physics from many resultants of black hole decay.
The candidates of new physics are lepton flavor 
and lepton (baryon) number violation. And since 
black holes can create only one top quark or one W boson, 
black hole decay at linear colliders gives the new method of
the precision measurements of top and W.

\vt

{\bf Note Added}

After the submittion of this letter, we learned from J.~D.~March-Russell
that their paper \cite{March-Russell} was the first to think about
black holes in the $\TEV$-scale gravity and to discuss many features,
say temperature and size for example.

And we also learned from K.~Cheung that the large entropy is the 
necessity to tell the object is truly a black hole ($S_{BH} > 25$),
and their papers \cite{Cheung} emphasized the fact. We calculated
the entropy of black holes in our setup. The result was
$S_{BH} = 31$, and we can say that our setup is enough to make
virtual black holes.

{\bf Acknowledgment}

Y.U. thank Japan Society for the Promotion of Science for financial
support.

\vt

\end{document}